\documentclass[twoside]{dis07}
\usepackage{graphicx,epsfig,color}

\def\beq{\begin{equation}}
\def\eeq{\end{equation}}
\def\beqa{\begin{eqnarray}}
\def\eeqa{\end{eqnarray}}
\def\as{\alpha_s}
\newcommand{\Tr}{\mathrm{Tr}\,}
\def\eq#1{Eq.~(\ref{#1})}

\begin{document}
\title{Power corrections for jets at hadron colliders}

\author{Matteo Cacciari$^1$, Mrinal Dasgupta$^2$, Lorenzo Magnea$^3$ and Gavin Salam$^1$
%
%
\vspace{.3cm}\\
%
1- LPTHE, CNRS UMR 7589, Universit\'e P. et M. Curie (Paris 6) \\
Universit\'e D. Diderot (Paris 7), 75252 Paris Cedex 05, France
%
\vspace{.1cm}\\
2- School of Physics and Astronomy, University of Manchester\\
Oxford Road, Manchester M13 9PL, U.K.
\vspace{.1cm}\\
3- Dipartimento di Fisica Teorica, Universit{\`a} di Torino, and\\
INFN, Sezione di Torino, Via P. Giuria, I--10125 Torino, Italy\\
}

\maketitle

\begin{abstract}
We discuss non-perturbative QCD corrections to jet distributions in hadron
collisions, focussing on hadronisation and underlying event contributions.
Using soft gluon resummation and Monte-Carlo modelling we show that hadronisation dominates at small values  of the jet radius $R$, behaving 
as $1/R$, while underlying event corrections grow with the jet area. This 
provides a handle to disentangle them and parametrize them in terms of measurable QCD parameters, which might enjoy a degree of universality.
\end{abstract}

\section{Introduction}
\label{intr}

With the advent of the LHC particle physics will once again break into new 
territory at the high energy frontier. With $14$ TeV available in the center of
mass, one would naively expect that the dynamics of confinement and low-energy
QCD would decouple and, thanks to factorization, have a minimal influence on 
high-$p_\perp$ observables (a typical figure of merit being $\Lambda/p_\perp
\sim {\cal O} (10^{-3})$). In general, this expectation is not fulfilled. Even at 
very high energy, for example, no hadronic cross section can be precisely
determined without a detailed knowledge of parton distributions in the colliding
hadrons. Furthermore, a wide range of observables of interest for both Standard
Model and BSM physics must rely upon a definition of hadronic jets  and a
measurement of the jet energy scale. Experience gained at the Tevatron 
\cite{Bhatti:2005ai} shows, for example, that a $1 \%$ uncertainty in the jet
energy scale causes a $1$ GeV uncertainty in the top quark mass determination,
and is reflected in a $10 \%$ uncertainty in the single inclusive jet $p_\perp$
distribution at $p_\perp \sim 500$ GeV\footnote{For an early study of the impact
of power corrections on jet distributions, see \cite{Mangano:1999sz}.}. 

Assuming a jet has been reconstructed with an infrared and collinear safe 
algorithm, dependent upon a parameter $R$ defining its size in the 
rapidity-azimuth plane, the energy of the jet will differ from the energy of the 
hard parton that originated it because of a variety of physical phenomena. 
Radiation from the underlying event and from pileup will spill inside the jet 
cone, increasing its measured energy; on the other hand, radiation produced 
during hadronisation will spill outside the jet cone, leading to a negative 
correction. It is important to realize that, while some of these corrections 
are definitely outside the reach of perturbative calculations (this is certainly 
the case for pileup, and to a certain extent for the underlying event),
hadronisation corrections can be explored with perturbative methods, supplemented
by soft gluon resummation and power correction technology. These methods have
been successfully applied to simpler processes such as $e^+ e^-$ annihilation and
DIS (for a review, see~\cite{Dasgupta:2003iq}), where studies of event shapes
showed that power corrections to distributions can be efficiently parametrized in
terms of a limited number of non-perturbative parameters, enjoying a remarkable
degree of universality \cite{Berger:2003pk,Berger:2004xf}. Here we will apply for
the first time these methods to jet distributions in hadron collisions, a much more
challenging environment. We consider, as an example, the single inclusive jet
$p_\perp$ distribution: we show that perturbative methods lead to a prediction
for the $R$ dependence of the leading power correction, which turns out to be
singular (behaving as $1/R$) for hadronisation corrections, while, as might be
expected, it grows as $R^2$ for the underlying event. We go on to compare the
analytic result to Monte Carlo models, finding broad agreement as well as some
interesting differences in the details.

\section{Issues in soft gluon resummation for jet distributions}
\label{resu}

Soft gluon resummation for the single-inclusive jet $p_\perp$ distribution was
first performed in~\cite{Kidonakis:2000gi}, using techniques developed in~
\cite{Laenen:1998qw} and ~\cite{Kidonakis:1998bk}\footnote{A refinement of 
the implementation of Ref.~\cite{Kidonakis:2000gi} was recently proposed in 
\cite{deFlorian:2007fv}}. Since jet production in hadron collisions generically
involves at least four colored partons, this is the first case in which nontrivial 
mixing of colour structures occurs. Formally, the structure of the resummed 
cross section is
\beq
      E_J \displaystyle\frac{d^3 \sigma}{d^3 p_J} = \displaystyle\frac{1}{s}
      \exp \left[ \sum_{p = 1}^2{\cal E}_{\rm IN}^{(p)} + 
      \sum_{p = J,R}{\cal E}_{\rm OUT}^{(p)} \right] \cdot 
      \Tr \left[ H S \right]~. 
\label{sinclu}
\eeq
The factors ${\cal E}_{\rm IN}$ and ${\cal E}_{\rm OUT}$ exponentiate collinear
logarithms associated with initial state radiation of the colliding partons, and with
the measured and the recoil outgoing jets, respectively.  At the level of power
corrections, these factors are expected to generate contributions of order 
$(\Lambda/p_\perp)^2$, associated with jet mass effects. These corrections 
are negligible, and therefore  we will concentrate on the contributions of soft
gluons emitted at wide angles  from the jet, embodied in the last factor in 
\eq{sinclu}. Here the trace is taken in the space of representations of the color
group that can be constructed out of  the scattering hard partons, while $H$ 
and $S$ are matrices containing hard and soft gluon contributions respectively. 
The exponentiation of soft radiation and the structure of color mixing can be 
simply understood~\cite{Dokshitzer:2005ig} by resorting to the eikonal
approximation. One can show that all soft logarithms can be organized in terms 
of eikonal colored dipoles, given by
\beq
{\cal D}_{ij} (Q, Q_0) \equiv \int_{Q_0}^Q d \kappa_\perp^{(i j)} 
\kappa_\perp^{(i j)} \as \left( \kappa_\perp^{(i j)}  \right) \int 
\, d \eta \, \frac{d \phi}{2 \pi} \, \frac{p_i \cdot p_j}{p_i 
\cdot k \, p_j \cdot k} \, T_i \cdot T_j~,
\label{logdip}
\eeq
where the indices $i,j$ label hard partons, $T_i$, $T_j$ are the color 
generators in the corresponding representation, and $\kappa_\perp^{(ij)} = 
2 \, p_i \cdot k \, p_j \cdot k/(p_i \cdot p_j)$ is the transverse momentum 
with respect to the emitting dipole. For $n < 4$ hard partons, all products of 
color matrices can be expressed in terms of Casimir operators, and thus there 
is no mixing of color structures.

Before extending the discussion to power corrections, it should be noted that
usage of resummations such as \eq{sinclu} requires great care both in the 
choice and in the definition of the observable. In general jet cross sections, 
which involve an explicit slicing of phase space, are affected by nonglobal
logarithms \cite{Dasgupta:2001sh}. In the present case, it turns out that 
the $p_\perp$ distribution is a global observable. Nonglobal logarithms will
generically spoil \eq{sinclu} at NLL level, and might influence power correction 
in a way which is not currently understood. One should also note that the 
choice of jet algorithm may affect the results quite drastically: IR safety is a 
must; furthermore, we will work assuming that the jet momentum is 
reconstructed using four-momentum recombination.

\section{Radius dependence of power corrections}
\label{poco}

In order to analyze the structure of power corrections using the resummation, 
we consider separately each dipole in \eq{logdip}. For the sake of simplicity 
we place the measured jet at zero rapidity, which does not qualitatively affect 
our results. Let $\delta \xi^{\pm} \left(k_\perp, \eta, \phi \right)$ be the
contribution of a soft gluon with momentum $k$ to the observable, which we
normalize as $\xi \equiv 1 - p_\perp/\sqrt{S}$. Note that the contribution is
different if the gluon is recombined with the jet ($\delta \xi^+$), or is left out 
of it ($\delta \xi^-$). We then construct the shift in the $\xi$ distribution due
 to the $(i j)$ dipole by integrating $\delta \xi^{\pm} \left(k_\perp, \eta, \phi \right)$ over the gluon phase space, with a measure given by the eikonal 
dipole, 
\beq
  \Delta \xi_{i j}^\pm (R) \equiv \int_\pm d \eta \frac{d \phi}{2 \pi} 
  \int_0^{\mu_F} d \kappa_\perp^{(ij)} \, \as \left( \kappa_\perp^{(ij)} \right) 
  k_\perp \left| \frac{\partial k_\perp}{\partial \kappa_\perp^{(ij)}} \right| \,
  \frac{p_i \cdot p_j}{p_i \cdot k \, p_j \cdot k} \, 
  \delta \xi^{\pm} \left(k_\perp, \eta,\phi \right)~.
\label{masterop}
\eeq
Note that $\delta \xi^+$ is integrated inside the jet cone, and $\delta \xi^-$ outside of it. The result is then a function of the jet radius $R$.

Different dipoles give different $R$ dependences, with a transparent physical
interpretation. The dipole constructed out of the two incoming partons has a
leading power correction growing like $R^2$, and it is natural to interpret it
as the way in which the buildup of the underlying event begins to be seen from
perturbation theory. Dipoles involving the measured jet, on the other hand, behave
as $1/R$ for small R. This behavior arises from gluons which are not recombined
with the jet: at small $R$, they are allowed to be emitted very close in phase
space to the radiating parton, and they begin to see the corresponding collinear
singularity. All dipole integrals are expressed in terms of a low-energy moment of
the strong coupling, ${\cal A} (\mu_F)$, which in principle could be related to
the analogous quantity measured in $e^+ e^-$ annihilation. As an example, we
consider the contribution of the $ q q  \rightarrow q q$ channel to the $p_\perp$
distribution. The dipole comprising the two incoming quarks gives
\beq
\Delta \xi_{{\rm in-in}} (R)  =  - \frac{4}{\sqrt{S}} \, {\cal A} (\mu_F)  
\, R \, J_1 (R) = - \frac{1}{\sqrt{S}} \, {\cal A} (\mu_F)  \, 
\left( 2 R^2 - \frac{1}{4} R^4 + \mathcal{O} \left( R^6 \right) 
\right)~.
\label{fin12}
\eeq
The dipole involving an incoming leg and the measured jet, on the other hand, 
gives
\beq
\Delta \xi_{\rm in-J} (R) = \frac{1}{\sqrt{S}} \, {\cal A} (\mu_F) 
\left(\frac{4}{R} - \frac{5}{4} R + \frac{23}{768} R^3 + \mathcal{O} 
\left( R^5 \right) \right)~.
\label{finfin1J}
\eeq
Notice that, as expected, this corresponds to a {\it negative} shift in the
$p_\perp$ distribution, given the definition of $\xi$. We remark also that
all singular contributions are essentially abelian in nature and can be directly collected into an overall shift in the physical distribution, weighed by $C_F$
($C_A$) for a quark- (gluon-) initiated jet. Terms subleading in $R$, however,
including \eq{fin12}, must be recombined taking color mixing into account,
and their contribution is more intricate than a simple shift.

\section{Monte-Carlo analysis of hadronisation and underlying event}
\label{mchu}

A powerful cross-check of the validity of the above discussion is
provided by examining hadronisation corrections for a range of
jet algorithms in Monte Carlo simulations. We consider dijet events
from both Pythia~\cite{Pythia} and Herwig~\cite{Herwig} and select
events whose highest-$p_\perp$ jet at parton-level has $55 < p_\perp
< 70$~GeV. The selection at parton level is intended to eliminate
selection bias associated with non-perturbative effects, as is
appropriate for comparison with our analytical formulae. 
For each event we examine the average difference in $p_t$ for the two
leading jets at hadron-level as compared to parton level. This difference is
separated into a hadronisation component and an underlying event contribution, 
the latter `defined' as that obtained when switching on underlying event 
(UE) and/or multiple-interactions in each Monte Carlo program, using the default
parameters of each. The UE contribution is expected to be uniform
in rapidity and azimuth, scaling as $\pi R^2$ for small $R$, or as $2 \pi R 
J_1(R)$ for general $R$ with $E$-scheme (4-momentum) recombination.

\begin{figure}
\centerline{\includegraphics[height=0.95\textwidth,angle=270]{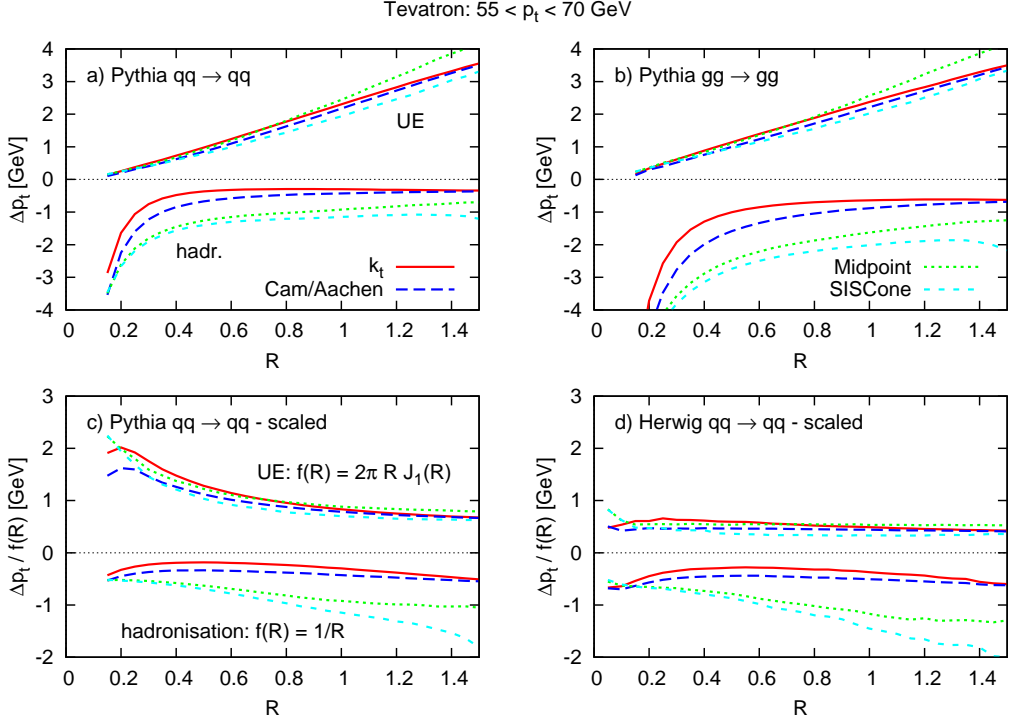}}
\caption{Shifts in $p_t$ associated with hadronisation and
  underlying-event contributions for the two leading jets in Pythia
  6.410 and Herwig 6.510 dijet events. 
}
\label{Fig:MV}
\end{figure}

The upper plots of Fig.~\ref{Fig:MV} show the two non-perturbative
components, for two hard scattering channels, as a function of $R$ for
the $k_t$ \cite{kt}, Cambridge/Aachen~\cite{CA}, Midpoint\footnote{This 
algorithm is infrared unsafe and should be thought of as a ``legacy'' algorithm,
shown only for historical purposes.} cone~\cite{Blazey} and
SISCone~\cite{SISCone} jet algorithms (all run through FastJet~\cite{FastJet}). 
The negative result for hadronisation, divergent at small $R$, is consistent with
\eq{finfin1J}. The rough factor of two between the hadronisation corrections 
for the $qq\to qq$ and $gg \to gg$ is consistent with the $C_A/C_F$ 
ratio expected from Sect.~\ref{poco}. One notes that the cone-type algorithms
have more negative corrections, differing from the sequential recombination
algorithms by a term roughly independent of $R$ whose explanation is
beyond the scope of the single-gluon calculation given above.
The UE event contribution to the jet $p_\perp$ is positive, as expected,
and for the typical range of $R$ studied experimentally, $0.4<R<0.7$,
similar in magnitude to the hadronisation correction. It is largely
identical for all algorithms, and roughly independent of the hard
scattering channel.

The analytical $R$ dependence can be studied in more depth by scaling
out the expectations for hadronisation and UE, as is shown in the
lower plots of Fig.~\ref{Fig:MV}. For recombination algorithms the result 
is roughly independent of $R$, as expected from \eq{finfin1J}. The 
normalisation of $\sim 0.5$~GeV is consistent with the magnitude of 
hadronisation corrections extracted at LEP. In contrast, the normalised UE correction is not constant in $R$. This can be shown to be a consequence of Pythia's implementation of colour-reconnections between the hard partons and
the underlying event and when examining models without such
reconnections, such as Herwig (and also Jimmy~\cite{Jimmy}),
Fig.~\ref{Fig:MV}d, this effect disappears. One notes that if one
extracts the UE $p_\perp$ density per unit rapidity (in the figure it is
normalised per unit area) then it is $\sim 6$ times larger than the
similarly normalised hadronisation correction, suggestive of far more
violent non-perturbative dynamics.

\section*{Acknowledgements}

We are grateful to Torbjorn Sj\"ostrand, Peter Skands and Jon
Butterworth for assistance and helpful comments. Work supported in
part by the EU Research and Training Network `HEPTOOLS' under contract
MRTN-CT-2006-035505 and by grant ANR-05-JCJC-0046-01 from the French
Agence Nationale de la Recherche.

\begin{footnotesize}

\end{footnotesize}

\end{document}